\documentclass[pra,aps,floatfix,twocolumn,showpacs,reprint,superscriptaddress]{revtex4}

\usepackage[T1]{fontenc}             
\usepackage{times}                
\usepackage[scaled=.90]{helvet}
\usepackage{graphicx}                
\usepackage{setspace}                
\usepackage[sort&compress]{natbib}   
\usepackage{float}
\usepackage[loose,nice]{units}       
\usepackage{tabularx}                
\usepackage{booktabs}                
\usepackage{amsmath}

\newcommand{\innerprod}[2]{\langle #1 | #2 \rangle}
\newcommand{\cg}[6]{\innerprod{#1 #2 #3 #4}{#5 #6}}

\newcommand{\pdiff}[2]{\ensuremath{\frac{\partial#2 }{\partial#1}}}
\newcommand{\pdifforder}[3]{\ensuremath{\frac{\partial^{#2}#3}{\partial{#1}^{#2}}}}

\newcommand{\matr}[1]{\ensuremath{\mathbf{#1}}}
\newcommand{\vect}[1]{\ensuremath{\mathbf{#1}}}

\newcommand{\au}{a.u.}
\newcommand{\etal}{\textit{et al.}}

\newcolumntype{C}{>{\centering\arraybackslash}X}
\newcolumntype{L}{>{\raggedleft\arraybackslash}X}
\newcolumntype{R}{>{\raggedright\arraybackslash}X}

\begin{document}

\title{A numerical study of two-photon ionization of helium using the Pyprop framework}

\author{R. Nepstad}
\email{raymond.nepstad@ift.uib.no}
\affiliation{Department of Physics and Technology, University of Bergen, N-5007 Bergen,
Norway}

\author{T. Birkeland}
\email{tore.birkeland@math.uib.no}
\affiliation{Department of Mathematics, University of Bergen, N-5007 Bergen, Norway}

\author{M. F\o rre}
\email{morten.forre@ift.uib.no}
\affiliation{Department of Physics and Technology, University of Bergen, N-5007 Bergen,
Norway}

\begin{abstract}
Few-photon induced breakup of helium is studied using a newly developed
\textit{ab initio} numerical framework for solving the six-dimensional
time-dependent Schr\" odinger equation. We present details of the method and
calculate (generalized) cross sections for the process of two-photon
nonsequential (direct) double ionization at photon energies ranging from 39.4 to
\unit[54.4]{eV}, a process that has been very much debated in recent years and is
not yet fully understood. In particular, we have studied the convergence
property of the total cross section in the vicinity of the upper threshold
($\sim$\unit[54.4]{eV}), versus the pulse duration of the applied laser field. We find
that the cross section exhibits an increasing trend near the threshold, as has
also been observed by others, and show that this rise cannot solely be
attributed to an unintended inclusion of the sequential two-photon double 
ionization process, caused by the bandwidth of the applied field.
\end{abstract}


\maketitle


\section{Introduction}
The study of how light interacts with matter has occupied physicists for a long
time. Of particular fundamental interest is single and multiple ionization of
atoms and molecules by photon impact, with subsequent ejection of one or a
multiple number of electrons. In this respect, single-photon multiple ionization
is special, as exchange of energy between the involved electrons is a
prerequisite for the process to take place. The investigation of such correlated
dynamical processes poses many unique challenges to experiment and theory. A
prime example of this is the one-photon double ionization of helium, which has
been the subject of intense study since the pioneering work of Byron and
Joachain~\cite{Joachain1967}. As a matter of fact, it was only recently that a
complete agreement between theoretical calculations and accurate measurements
with synchrotron radiation was established, for the value of the (generalized)
cross section for the direct (nonsequential) double ionization
process~\cite{Briggs,Huetz,Samson1998,Foumouo2006}.

The problem of two-photon double ionization of two-electron atomic systems
presents additional difficulties. First, the separation between sequential and
nonsequential double ionization often becomes a nontrivial
problem~\cite{Lambropoulos_PRA_2008}, in situations where both processes are
energetically accessible~\cite{FeistPRL2009}. Here, 'sequential' ionization
usually refers to the process where the electrons are emitted one after the
other by subsequent absorption of a photon each, and where the second electron
has time to relax to a stationary ionic state before it is emitted. Thus, energy
exchange between the two electrons is not strictly required. In contrast,
'nonsequential' double ionization depends upon exchange of energy between the
outgoing electrons, and as such it represents a clear departure from an
independent-particle picture. As mentioned above, after more than 40 years of
investigation, the case of double electron ionization by single photon absorption
is now considered to be a well understood process~\cite{Schneider2002}, but the
related problem of two-photon direct (nonsequential) double ionization in helium,
at photon energies ranging from 39.4 to $\unit[54.4]{eV}$, is still being
debated~\cite{Pindzola,Feng2003, Bachau2003,Bachau_EPD,Hu2005,Foumouo2006,
Shakeshaft2007, Ivanov2007, Horner2007, Nikolopoulos2007,
Feist2008,Guan2008,Foumouo2008,Palacios2009,Rudenko}.  In particular, the
separation of two-photon single ionization, where the remaining electron is left
in an (excited) ionic state, from two-photon double ionization, has turned out
to be a subtle theoretical problem, as the role of electron correlations in the
final states is not yet fully understood~\cite{Foumouo2006}. Moreover, moving
beyond the single-photon ionization regime is extremely challenging from the
experimental point of view, due to the weakness of the signals. Although the
total cross section for the two-photon nonsequential double ionization of helium
has been the subject of experiments, employing state of the art high-order
harmonic~\cite{Hasegawa_PRA_2005,Nabekawa} and free-electron laser (FEL)
radiation~\cite{Sorokin2007}, the experimental results remain
inconclusive~\cite{Antoine}.

In this paper, we revisit the problem of two-photon nonsequential double
ionization in helium.  We present details on a recently developed B-spline based
numerical framework for solving  the two-electron time-dependent Schr\"odinger
equation. The method is built on the more general Pyprop framework~\cite{pyprop}
and was recently used to study the role of electron correlations in the
stabilization of helium in intense extreme-ultraviolet (XUV) laser
fields~\cite{Torebi_PRL}.  The two-electron module we have implemented is
designed to utilize massively parallel supercomputers to perform accurate
large-scale time-dependent simulations, and we here use it to make a
contribution to the ongoing discussion on two-photon double ionization.

For the total cross section we obtain values that are in close agreement with
some recently reported results~\cite{Feist2008,Palacios2009}.  In contrast, our
results are in clear disagreement with results from calculations where
correlation effects in the final scattering states have been included to some
extent, using different approximative methods. In our approach, which is similar
to the method adapted by Feist {\it et al.}~\cite{Feist2008}, and others, this
problem is circumvented by letting the ionized wave packet propagate for some
time after the end of the laser pulse, so that most of the wave packet
eventually reaches the asymptotic (Coulomb) region~\cite{Madsen2007}.  The wave
function is then analyzed by means of projections onto a set of uncorrelated
eigenstates, i.e., Slater determinants constructed from one-electron orbitals. 

With our present method we are able to consider relatively long pulses (up to
about $\unit[10]{fs}$ in total pulse duration). This puts us in a position to
study the convergence property of the total cross section in the vicinity of the
upper threshold ($\sim\unit[54.4]{eV}$), as a function of pulse length.  It has
been reported that the cross section exhibits an apparent sharp rise near the
threshold~\cite{Horner2007,Shakeshaft2007,Horner2008,Feist2008,Palacios2009}.
This rise stands somewhat in contrast to the results obtained in lowest
(nonvanishing) order perturbation theory (LOPT)~\cite{Lambropoulos_PRA_2008}.
Our calculations with longer pulses indicate that the cross section indeed
appears to grow as it approaches the threshold (up to $\unit[51.7]{eV}$), but we
cannot rule out the possibility that the cross section reaches a maximum at some
point in the immediate neighborhood of (or on) the threshold.  Based on our
results we are tempted to conclude that the increase of the cross section around
$\unit[54.4]{eV}$ is not solely due to an unintended inclusion of the sequential
process~\cite{Lambropoulos_PRA_2008,Horner2008}.  If this is correct, the very
interesting question remains: what is the underlying physical process causing
the unexpected behavior?

The rest of this paper is organized into two main sections. The first outlines
some theoretical background, and then goes on to discuss the various aspects of
the numerical method, ending with a discussion of convergence issues.  In the
final section, we present our calculations of the total cross section, and pay
particular attention to its behavior near the sequential threshold.

Atomic units ($\hbar$, $m_e$ and $e$ replaced by $1$) are used in the following
exposition, except where otherwise noted.

\section{Theory and numerical approach}
B-splines~\cite{Boor1978} have long been popular in atomic and molecular
physics computations, due to their ability to accurately represent atomic
eigenstates (see~\cite{Bachau2001} and references therein). For time-dependent
calculations, integration of the Schr\"odinger equation directly in the
B-spline representation is very efficient for one-electron systems, due to the
sparse and structured matrices that arise~\cite{Cormier1997}. For two-electron
systems, however, the matrix structure becomes more complicated, and the matrix
sizes are also much larger. Therefore, in time-dependent approaches for
two-electron systems using B-spline discretization, one-electron orbitals or
atomic eigenstates have been used to construct a matrix representation of the
field interactions. Both these approached are useful and accurate, but do not
easily scale to very large basis sizes, because the basis functions are global,
resulting in dense matrices. Operations involving such matrices are difficult
to parallelize efficiently, which eventually becomes necessary as the basis
size is increased.

In this section, we present some details of a recently developed B-spline based
numerical approach to the two-electron time-dependent Schr\"odinger equation,
where the time integration is performed directly in the B-spline basis. A
Python/C++ implementation of the method have been created, which uses the
Pyprop framework~\cite{pyprop}.

%
%
\subsection{Theoretical background}
We consider the two-electron time-dependent Schr\"odinger equation (TDSE),
\begin{equation}
	i\frac{\partial}{\partial t}\Psi(\vect{r}_1,\vect{r}_2,t)=H\Psi(\vect{r}_1,\vect{r}_2,t).
\end{equation}
Employing the semi-classical approximation, the light-atom interaction Hamiltonian can be cast into
the form,
\begin{eqnarray}
    H &=&\left(\frac{\vect{p}_1^2}{2} - \frac{2}{r_1}
	+ H_{f,1}\right) + \left(\frac{\vect{p}_2^2}{2} - \frac{2}{r_2} +
	H_{f,2}\right)\nonumber\\ &&+ \frac{1}{|\vect{r}_1 - \vect{r}_2|},
	\label{eq:heliumham} 
\end{eqnarray}
where $H_{f,i}$ represents the interaction with the external field.
The laser-atom interaction is modeled in the dipole approximation using the
velocity gauge formulation, which, when linearly polarized along the $z$ axis,
can be written as
\begin{equation}
    H_{f,i} = A_z(t) p_{z_{i}}.
\end{equation}
Here $A_z$ is the vector potential defining the external field.  The
corresponding electric field is given by $E_z=-\partial A_z/\partial t$.  For
the temporal form of the laser interaction, a sine-squared carrier-envelope was
chosen, i.e.,
\begin{equation} 
	A_z(t) =A_0 \sin^2\left(\frac{\pi t}{T}\right)\cos(\omega t), 
\end{equation}
where $A_0=E_0/\omega$, $E_0$ being the electric field amplitude, $\omega$ is
the (central) laser frequency, and $T$ defines the (total) pulse duration. We
have also considered pulses with a Gaussian envelope,
\begin{equation}
    A_z(t) =
    A_0\exp\left[-2\ln{2}\left(\frac{t-t_0}{\tau_0}\right)^2\right]\cos{\left[\omega(t -
    t_0)\right]},
    \label{eq:gausspulse}
\end{equation}
where $\tau_0$ is the full-width at half-maximum (FWHM) pulse duration, 
and $T=2t_0$ defines the (chosen) total pulse duration.

%
%
\subsection{Discretization} 
Turning the continuous TDSE into a set of coupled ordinary differential
equations is achieved by representing the wave function, in spherical
coordinates, as a product of radial B-splines and Coupled Spherical Harmonics,
\begin{equation} 
    \psi(\vect{r}_1, \vect{r}_2, t) = \sum_{i,j,k}c_{ijk}(t)\frac{B_i(r_1)}{r_1}\frac{B_j(r_2)}{r_2}
	\mathcal{Y}^{LM}_{l_1,l_2}(\Omega_1, \Omega_2).  
\end{equation} 
Here, $k = \{ L, M, l_1, l_2 \}$ is a combined index for the angular indices,
and the Coupled Spherical Harmonics are given in terms of Spherical Harmonics
as~\cite{Bransden2003}
\begin{equation}
	\label{eqn:coupled-sph}
	\mathcal{Y}_{l_1 l_2}^{LM} = \sum_m \cg{l_1}{l_2}{m}{M-m}{L}{M} 
	Y_{l_1}^m(\Omega_1) Y_{l_2}^{M-m}(\Omega_2).
\end{equation}
For the special case of $z$-polarization, the problem reduces effectively to five
dimensions, as $M$ is conserved during the laser-atom interaction. Since we are
studying ionization from the ground state ($M = 0$ manifold), only $M=0$ is
included in the calculations.  The B-spline basis functions depend upon several
parameters, and determining the optimal values of these are not trivial (see
Bachau \etal~\cite{Bachau2001} for a discussion).  Throughout this paper, we
have used order 5 B-splines, and an exponential distribution of breakpoints
near the origin, with linear spacing further out, providing accurate
representation of both bound and continuum states. Zero boundary conditions are
imposed by removing the first and last B-spline for each radial direction.

Except for the electron-electron interaction, all the terms in Eq. 
(\ref{eq:heliumham}) are one-electron operators, and are straightforward to
discretize. When calculating matrix elements of these, the radial integrals are
performed with Gauss-Legendre quadrature~\cite{Bachau2001}, while the angular
integrals are handled analytically. The electron-electron interaction is first
expanded in a truncated multipole series,
\begin{equation}
    \frac{1}{\vect{r}_1 - \vect{r}_2} \approx
    \sum_{l=0}^{l_{max}}\ \sum_{m=-l}^{l}\frac{4\pi}{2l+1}\frac{r_<^l}{r_>^{l+1}}
	Y_{l,m}^*(\Omega_1)Y_{l,m}(\Omega_2),
\end{equation}
and each of the terms are handled in a similar manner to the one-electron
operators.  Finally, taking into account the non-orthogonality of the B-spline
basis, $\int dr  B_i(r) B_j(r) = S_{ij}$, the TDSE can be written in matrix
form
\begin{equation}
    \imath \matr{S} \pdiff{t}{} \vect{c}(t) = \matr{H}(t)\vect{c}(t).  
\end{equation}
The total overlap matrix $\matr{S}$ is a Kronecker product of one-electron
overlap matrices for each angular momentum component,
\begin{equation} 
    \matr{S} = \matr{I}_k \otimes \matr{S}_1 \otimes \matr{S}_2.  
\end{equation}
In contrast to the one-electron case, the Hamiltonian matrix $\matr{H}$ does not
have a simple banded structure, although it is quite sparse. An overall banded
structure remains, but is now interlaced with bands of zeros.

At present, we make use of multiprocessor systems by distributing the
wavefunction across several processor in the angular rank, such that all the
time-independent radial matrix elements are processor-local. This of course
restricts the number of processors we may use to the total number of angular
momentum elements. We are currently implementing the option of distributing one
or both radial ranks, which would allow us to use more processors, and thus an
even larger radial basis.

%
%
\subsection{Time integration} 

It is common for numerical integrations schemes to be based on the first-order
exponential approximation to the propagator,
\begin{equation} T(t_i, t_f) = \exp\left[-\imath\matr{S}^{-1}\matr{H}\Delta
	t\right] + \mathcal{O}(\Delta t^2), \end{equation}
which is quite accurate for reasonably small time steps $\Delta t = t_f - t_i$.
The matrix exponential may be calculated efficiently by the popular Arnoldi or
Lanczos methods~\cite{Park1986, Smyth1981, Feist2008}. However, instabilities
prevent us from using this approach in the present case; instead, we use the
implicit Cayley-Hamilton form of the propagation operator,
\begin{equation} 
    \label{eqn:cayley-hamilton}
    \left(\matr{S} + \frac{\imath\Delta t}{2}\matr{H}\right)\vect{c}(t+\Delta t) =
    \left(\matr{S} - \frac{\imath\Delta t}{2}\matr{H}\right)\vect{c}(t). 
\end{equation}
The above linear system of equations is typically too large to be solved by
direct methods, but very sparse. It is therefore solved using the iterative
Generalized Minimum Residual method (GMRES)~\cite{Saad1986, Saad2003}. Similarly
to the Arnoldi method, GMRES uses a Krylov subspace, constructed from successive
multiplications of $\left(\matr{S} + \frac{\imath\Delta t}{2}\matr{H}\right)$ on
$\vect{c}(t)$ in each time step. A least-square problem is solved in the
subspace spanned by these vectors, and a solution of the equation with a minimum
residual is obtained.  With this method, the error in the computed solutions
(residual) is controlled by the size of the Krylov subspace, which can be
increased automatically, thus always ensuring a high precision solution.

As is typical for discretized partial differential equations, and in particular
those obtained with a B-spline basis, the system is quite stiff, and a good
preconditioner is essential for the convergence of GMRES.
Let $\matr{A} = (\matr{S} + \frac{\imath\Delta t}{2}\matr{H})$ and $\vect{b} =
(\matr{S} - \frac{\imath\Delta t}{2}\matr{H})\vect{c}$. The preconditioner
$\matr{M}$ is then constructed to make the linear system 
\begin{equation}
	\matr{M}^{-1} \matr{A} \vect{c} = \matr{M}^{-1} \vect{b}
\end{equation}
easier to solve than the original system. This can be achieved by letting
$\matr{M}$ be an approximation of $\matr{A}$. The preconditioner used in this
paper is a block type preconditioner, where each block consists of the complete
radial Hamiltonian for a given coupled spherical harmonic ($k$ is the angular
index),
\begin{equation}
M_{(i,j,k), (i',j',k')} \!=\!\! \left(\!\!S_{(i,j,k),
	(i',j',k)} + \frac{\imath\Delta t}{2}H_{(i,j,k),
	(i',j',k)}\!\!\right)\!\! \delta_{k,k'}.
\end{equation}
This block diagonal matrix is distributed across processors in such a manner
that the elements in the wave function corresponding to one block are all local
to one processor. When solving linear systems involving $\matr{M}$, each block
can be solved separately and thus there is no communication between different
processors. Furthermore, as the exact solution of $\matr{M}$ is not required,
we employ the incomplete LU (ILU) factorization~\cite{Saad2003}, $\matr{M} \approx
\matr{L}\matr{U}$, as provided by the IFPACK toolkit available in the
Trilinos~\cite{Heroux2003} library. A similar preconditioner, using
SuperLU~\cite{Demmel1999} to solve $\matr{M}$ exactly, was also tested, but
found to be less efficient for the given system.

%
%
\subsection{Extracting physical information} 
Although excitation of the neutral atom is negligible for the intensities and
frequencies we will consider in this paper, calculation of a subset of
the eigenstates of the helium atom is nevertheless useful in many cases.  The
implicitly restarted Arnoldi method (IRAM) \cite{Sorensen1992} may be used for
this purpose. For reasons similar to those prompting the use of a
preconditioner above, IRAM will converge slowly for interior eigenvalues.
However, shifted inverse iterations can be used to accelerate the convergence
for eigenvalues near a given shift $\sigma$,
\begin{equation}
	\left(\matr{H} - \sigma \matr{S} \right)^{-1} \vect{c}_n =
	\frac{1}{E_n - \sigma} \vect{c}_n \matr{S}. 
	\label{eq:inverse_it}
\end{equation}
IRAM requires the multiplication of the operator matrix on a vector in order to
operate. For inverse iterations this corresponds to solving the linear system in
Eq.~(\ref{eq:inverse_it}), for the matrix $\matr{B} = (\matr{H} - \sigma
\matr{S})$. Note the similarity between \matr{A} and \matr{B}, the difference
being only the scalars in front of $\matr{H}$ and $\matr{S}$. The linear solver
used for the propagation can therefore also be used to find eigenvalues with
IRAM. Alternatively, a preconditioned Davidson method can be used. We found that
when our basis grows sufficiently large, the Block-Davidson
approach~\cite{Arbenz2005}, as implemented in the Trilinos package Anasazi,
performed favorably compared to the shift-invert Arnoldi method. In most cases,
either of these methods can be used to rapidly obtain eigenpairs in the vicinity
of any given shift value.

In order to extract double ionization probabilities from the wavefunction, we
must project onto some set of states which span this space. The exact double
continuum states are hard to find, as they require solving a scattering problem
for the full two-particle system. An approximation using non-correlated states,
obtained by solving a set of one-electron radial eigenvalue problems, is used
instead,
\begin{equation} 
	\left(-\frac{1}{2}\pdifforder{r}{2}{} + \frac{l(l+1)}{2r^2} -
	\frac{Z}{r}\right)R^Z_{n,l}(r) = E^Z_{n,l}R^Z_{n,l}(r).
\end{equation}
The double ionization continuum is represented by a product of He$^+$ ($Z=2$)
states, which, when expanded in the B-spline basis, we will denote
$\vect{b}_{n_1, n_2, k}$.
These states are not orthogonal to the bound states of the
atomic system, and consequently the projection of the final wave function on the
atomic bound states should be removed before the analysis is performed.
Furthermore, the approximated double continuum states used here are not
eigenstates of helium as the electron-electron correlation is ignored. The
wave packet must therefore be propagated after the interaction until it reaches
the Coulomb zone~\cite{Madsen2007}, where the electron-electron interaction is
negligible.

To calculate the double ionization probability, we project the final wave
function onto the non-correlated double continuum functions $\vect{b}$, and sum
over all contributions,
\begin{equation}
    P_\mathrm{double} =
    \sum_{n_1, n_2,k}\left|\vect{b}_{n_1, n_2, k}\cdot\vect{c}(T)\right|^2.
\end{equation}
Having found the double ionization probability, we may then determine the total
cross section for the nonsequential two-photon double 
ionization process~\cite{Madsen2000, Feist2008},
\begin{equation}
	\sigma_{DI} = \left(\frac{\omega}{I_0}\right)^2\frac{P_{\mathrm{double}}}{T_{eff}},
\end{equation}
where $I_0$ is the pulse intensity. The finite duration of the pulse is
accounted for by $T_{eff}$, which for a two-photon process
reads~\cite{Madsen2000, Feist2008}
\begin{equation}
    T_{eff} =
	\int_{-\infty}^\infty\limits{\left[\frac{I(t)}{I_0}\right]^2\mathrm{dt}}.
\end{equation}
For a sine-squared pulse envelope $T_{eff} = 35T/128$, while for a Gaussian
envelope $T_{eff,g} = \tau_0/2\sqrt{\pi/2\ln 2}$.

%
%
\subsection{Accuracy and numerical convergence} 
The reliability of numerically calculated quantities must be checked carefully,
and this is usually performed by varying the relevant numerical parameters and
studying the resulting changes.  It is possible, from certain physical
considerations, to obtain a reasonable \textit{a priori} estimate for the values
of some parameters. In other cases, simple numerical calculations may be
performed to get such estimates. For example, in the present case the number of
photons absorbed in the system determines the maximum value of the total angular
momentum quantum number $L$ that must be included in the basis, and also the
required radial box size, from an estimate of the energy of the ejected
electrons. The remaining radial basis parameters mainly determine the quality of
the ground state (see Table~\ref{Table1}), and the maximum photoelectron energy
supported.  By solving the related one-dimensional radial eigenvalue problem
(Coulombic potential, $Z=2$), estimates for these parameters can be obtained.
Specifically, estimating the density of states from $1/\Delta E_n$, $\Delta E_n$
being the energy separation between state number $n$ and $n+1$ in the
discretized basis, and comparing with the known value, we may determine to what
extent the density of states is correctly represented in the box, in the energy
region of interest.  Finally, by sampling the parameter space in the vicinity of
the values thus estimated, a good indication as to whether the results are
converged is obtained.

\heavyrulewidth=.08em \lightrulewidth=.05em \cmidrulewidth=.03em
\belowrulesep=.65ex \belowbottomsep=0pt \aboverulesep=.4ex \abovetopsep=0pt
\cmidrulesep=\doublerulesep \cmidrulekern=.5em \defaultaddspace=.5em

\begin{table}
\centering
\begin{tabularx}{.9\columnwidth}{RCC} 
	\toprule 
	Quantity & Calculated value
	(a.u.) & Reference value (a.u.)\\
	\midrule He ($1^1S$)  & -2.903 667 & -2.903\ 724\\ 
	He ($2^1S$) & -2.145\ 971 & -2.145\ 974\\ 
	He ($1^3S$) & -2.175\ 229 & -2.175 229\\ 
	H$^-$ ($1^1S$) & -0.527\ 735 & 0.527\ 751\\
	Li$^+$ ($1^1S$) & -7.279\ 827 & -7.279\ 913\\ 
	\bottomrule 
\end{tabularx}
\caption{Calculated energy levels in the helium atom and the helium-like
	ions H$^-$ and Li$^+$, compared with reference values from Drake~[2]. The
	calculations used $l_{max} = 7$ and 50 exponentially distributed B-splines
	for each radial direction, in a box extending to 50 a.u.}
\label{Table1}
\end{table}

Time step size must also be considered, since our implicit time integrator
incurs a local error of order $\Delta t^3$.  Incidentally, this error is one
order of magnitude smaller than that of the exponential propagator it
approximates, but the constants are typically different, and may depend on all
the other parameters except $t$. In any case, we have used a default time step
$\Delta t = \unit[0.01]{\au}$ Halving the time step to $\unit[0.005]{a.u.}$
produced changes in our calculated quantities of less than $0.1\%$.

\section{Results}

%
%
\subsection{One photon double ionization}
\begin{figure}[t] 
	\begin{center} 
		\includegraphics[width=8.5cm]{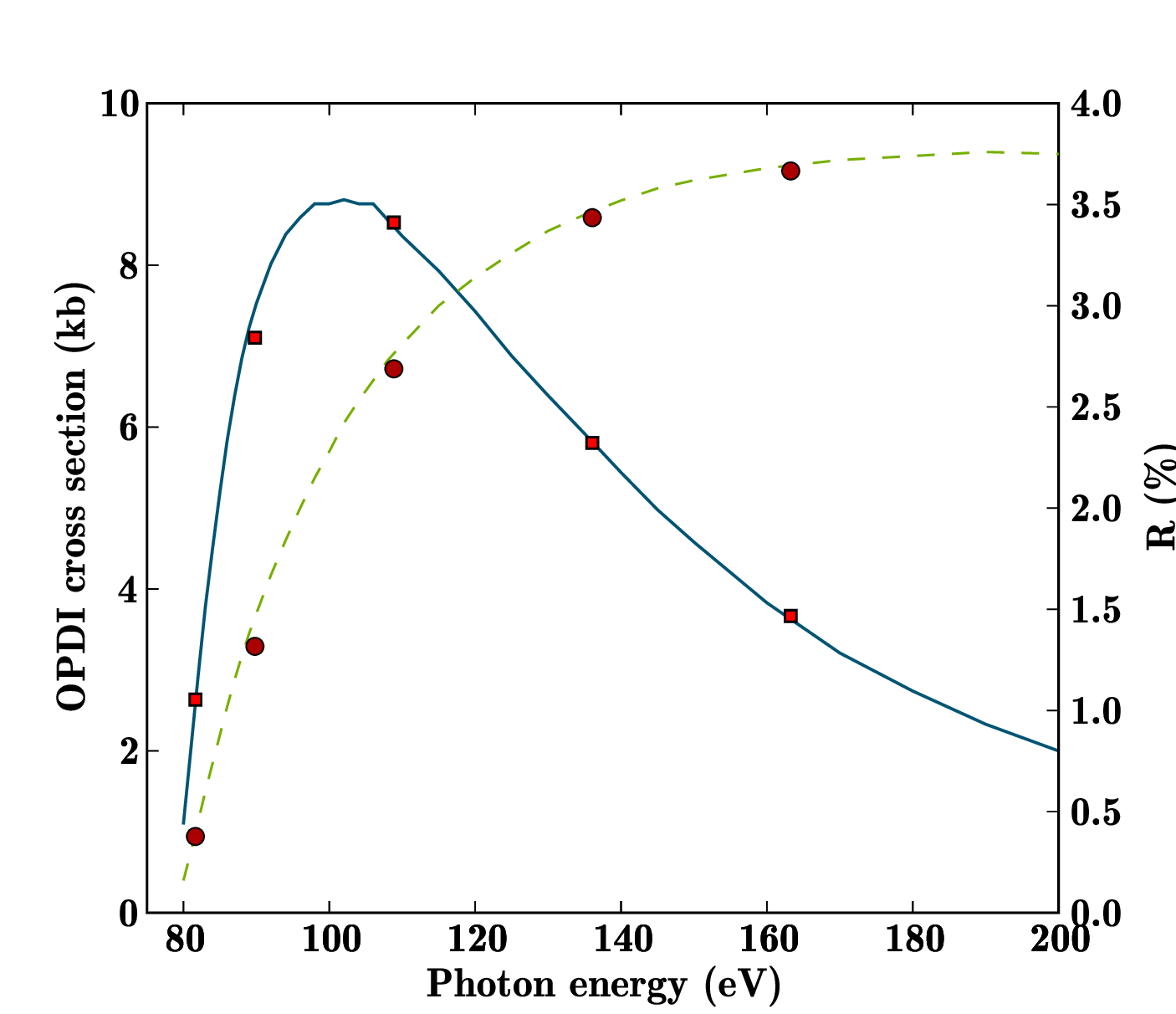} 
	\end{center}
	\caption{(Color online) A comparison of calculated one-photon double
	ionization cross sections (red squares) and experimental values from Samson
	\etal~\cite{Samson1998} (blue full line). The ratio of double to single
	ionization is also shown. Red dots; calculated result, and green dashed line; 
	experimental result.}
	\label{fig:opdi_crossection} 
\end{figure}

Because of the electron-electron interaction, double ionization of helium may
proceed through the absorption of a single photon. This one-photon double
ionization process has been investigated at length in both
theoretical~\cite{Schneider2002, Huetz, Briggs} and
experimental~\cite{Samson1998, Dorner1998, Bizau1995} studies, resulting in
close quantitative agreement in both total and differential ionization cross
sections, and a thorough understanding of the physical underpinnings. This makes
it an ideal benchmark against which new numerical schemes may be gauged.
Accordingly, we have calculated cross sections for selected photon energies in
the interval $\unit[80-200]{eV}$. A box with $r_{max} = \unit[160]{a.u.}$, 311
B-splines, $l_{max} = 5$, and values of $L_{max}$ up to five was used. The pulse
duration was set to 20 optical cycles, and the wave packet was propagated four
additional cycles after the pulse was over, allowing the ionized component to
reach the Coulomb zone. The results, shown in Fig.~\ref{fig:opdi_crossection},
include double ionization cross sections (red squares) and the ratio of double
to single ionization (red circles), both within $5\%$ of the experimental
results of Samson \etal~\cite{Samson1998}, who state the accuracy of their
results to be $\pm 2\%$. 

%
%
\subsection{Two photon double ionization}
\begin{figure}[t]
	\begin{center}
		\includegraphics[width=8.5cm]{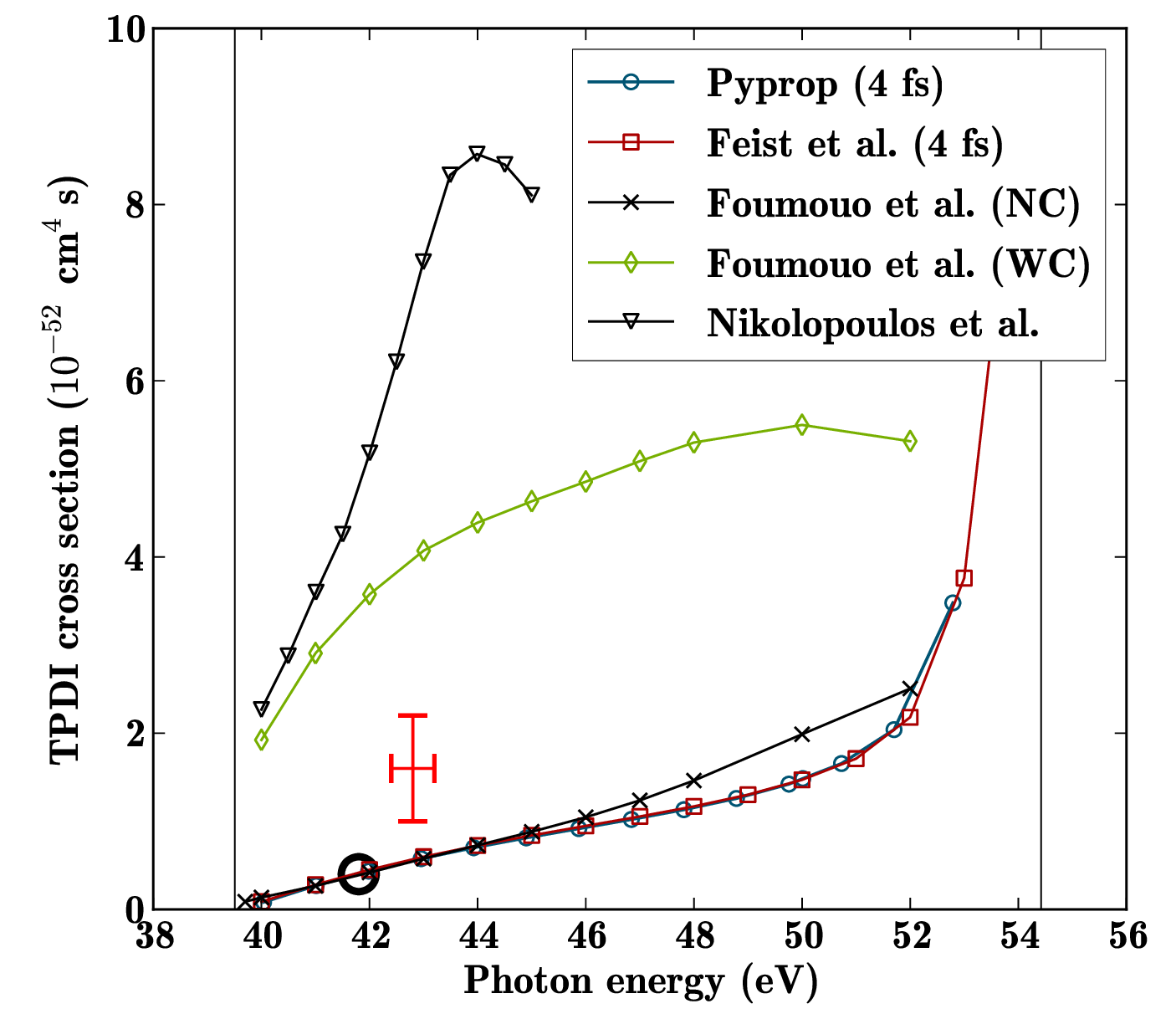}
	\end{center}
	\caption{(Color online) Two-photon double ionization cross sections. 
    Blue line with circles: the present results obtained with Pyprop,
    black circle: experimental result of Hasegawa
	\etal~\cite{Hasegawa_PRA_2005}, red cross: experimental result of Sorokin
	\etal~\cite{Sorokin2007}, red line with squares: Feist \etal~\cite{Feist2008}, green line
	with diamonds: Foumouo \etal~\cite{Foumouo2006} (with correlation, WC), black line with
	crosses: Foumouo \etal~\cite{Foumouo2006} (no correlation, NC), and black line with triangles: 
	Nikolopoulos~\etal~\cite{Nikolopoulos2007}. The
	vertical lines define the two-photon direct double ionization region.}
	\label{fig:tpdi_crossection}
\end{figure}

We now consider the problem of two-photon direct double ionization. Our results,
together with a small subset of results from the numerous studies available in
the literature, are shown in Fig.~\ref{fig:tpdi_crossection}. The calculations
have been made using a box extending to $r_{max} = \unit[250]{a.u.}$, and 246
B-splines, while the angular basis was truncated at $l_{max} = 5$ and $L_{max} =
3$, respectively. Additionally, the intensity of the laser field was fixed at
$\unit[10^{13}]{W/cm^2}$, which is well within the perturbative regime. We also
checked that decreasing the intensity by a factor of ten did not produce any
significant changes in the results; at $\omega = \unit[51.7]{eV}$ the change in
total cross section was less than $0.1\%$. Improving our basis by increasing
$(l_{1,max}, l_{2,max}, L_{max})$ to (7,7,5) and the number of B-splines to 270
resulted in only minor changes in the cross sections ($<0.03\%$ at
$\unit[42]{eV}$).  To facilitate the separation of single- and double continuum
components of the wave function, we ran the propagation algorithm an additional
femtosecond after the end of the pulse before performing the projections on the
Coulomb waves, to ensure that the major part of the wave packet had entered the
asymptotic Coulomb zone~\cite{Madsen2007,Feist2008}.

The agreement between our results and those of Feist \etal~\cite{Feist2008} is
particularly close, but this is not surprising due to the similarity of the
numerical methods and the projection method used to extract the double
ionization probability.  In contrast, the J-matrix result of Foumouo
\etal~\cite{Foumouo2006} (green line with diamonds in
Fig.~\ref{fig:tpdi_crossection}), and the perturbation theory result of
Nikolopoulos \etal~\cite{Nikolopoulos2007} (black line with triangles in
Fig.~\ref{fig:tpdi_crossection}), deviate significantly from ours, i.e., the
calculated total cross sections for the reaction differ by as much as an order
of magnitude.  In both of these approaches, correlation effects were included in
the final state, to some extent, while in our calculations no such effects were
included.  It should be noted, however, that Foumouo \etal~\cite{Foumouo2006}
obtained similar results to ours when they neglected completely the role of
electron-electron interactions in the final wave function (black line with
crosses in Fig.~\ref{fig:tpdi_crossection}).  This seems to stress the
importance of electron correlations in the final states, however, by propagating
the wave packet for a long time after the pulse, the correlation effect should,
in principle, be minimized, as argued and tested by Feist
\etal~\cite{Feist2008}. In that particular study, they employed a very large
grid and propagated the wave packet some $\unit[20]{fs}$ after the pulse, to
explicitly check for convergence of the cross section.  They also extracted the
double ionization directly from the grid representation, by partitioning the
radial grid, and by varying the partition limits found an upper bound for the
possible value of the cross section $\sim 25\%$ higher than their quoted results
(Fig.~\ref{fig:tpdi_crossection}).

Regarding the apparent rapid rise in the value of the calculated  cross section
near the sequential ionization threshold $(\unit[54.5]{eV})$, this is usually
attributed to the bandwidth of the pulses used in time-dependent methods and an
unwanted inclusion of the sequential
process~\cite{Horner2007,Horner2008,Lambropoulos_PRA_2008,Feist2008,Palacios2009}.
Thus, due to the finite spectral width of the pulse, the sequential process
cannot be completely separated from the nonsequential one, even below threshold.
Now, since the relative importance of the sequential process scales as $T^2$, as
opposed to the $T$-dependence  of the nonsequential, attempting to extract a
cross section when both processes are present would result in a divergent
behavior.  This problem can be circumvented by simply increasing the pulse
duration in order to lower its bandwidth. Following this procedure, one avoids
significant contribution from the sequential component up to some finite
distance from the upper threshold, but at a certain point the overlap with the
sequential region again becomes non-negligible, and the $T^2$ scaling law causes
an even sharper rise, due to the now longer pulse duration.  Examining the
relative importance of the different spectral components in the laser pulse as
the pulse duration is increased, one can show that the relative contribution
from the sequential process will ultimately become negligible, despite the
$T^2$-dependence of the ionization yield.  Thus, using successively longer
pulses, one may, at least in principle, resolve the behavior of the direct
two-photon double ionization cross section arbitrarily close to the threshold,
without contamination from the sequential process.

\begin{figure}[t]
	\begin{center}
		\includegraphics[width=8.5cm]{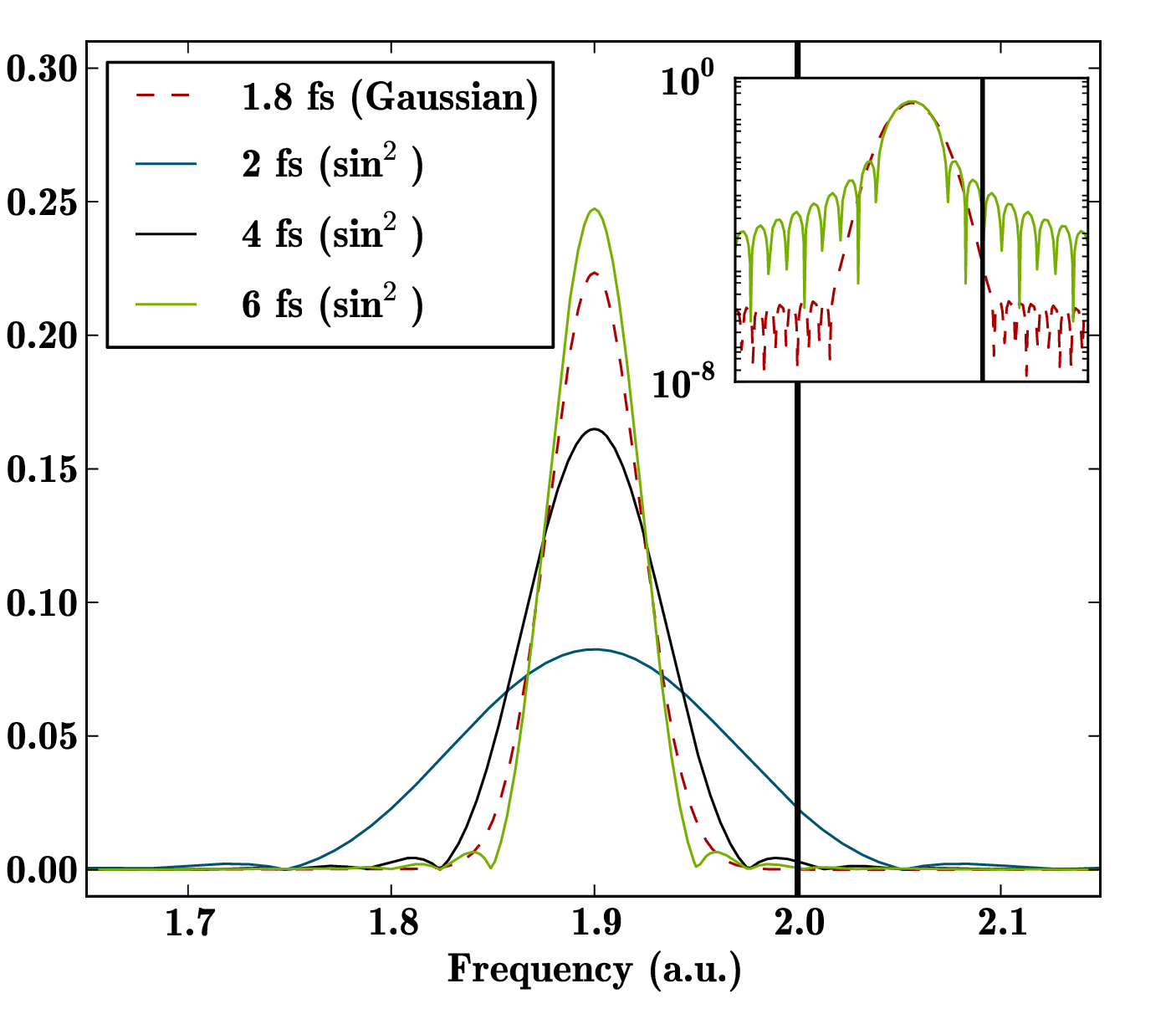}
	\end{center}
	\caption{(Color online) Fourier spectra of pulses with different temporal shapes and
    durations. Full lines represent sine-squared pulses with 
    $T=\unit[2]{fs}$ (blue), $\unit[4]{fs}$ (black) and $\unit[6]{fs}$ (green), 
    while the red dashed line
represents a Gaussian pulse with $\tau_0 = \unit[1.8]{fs}$ and $T=\unit[9.4]{fs}$. 
	The inset shows the $\unit[6]{fs}$ and Gaussian pulse spectra on a logarithmic scale.} 
	\label{fig:pulse_spectrum} 
\end{figure}

Pursuing this line of thought, we have performed some additional calculations at
$\omega = \unit[1.9]{a.u.}$ $(\unit[51.7]{eV})$ with longer pulse durations and
different temporal shapes, whose Fourier spectra are shown in
Fig.~\ref{fig:pulse_spectrum}. The $\unit[2]{fs}$ pulse has a clear extension
into the sequential region, indicated by the black vertical line, while the
longer pulses have successively less overlap.  The resulting cross section
values are shown in Table~\ref{tbl:cross_section}. For pulse durations in the
region $\unit[2-6]{fs}$, and a sine-squared envelope, we found that the cross
section leveled out at $\unit[2.0\times 10^{-52}]{cm^4s}$ as $T$ increased.
Changing the temporal profile of the pulse to a Gaussian one, cf.
Eq.~\ref{eq:gausspulse}, with a (total) duration of $\sim \unit[9.4]{fs}$, the
same value for the cross section was obtained.  In order to extract the cross
section at photon energies exceeding 52 eV, significantly longer pulse durations
than 5 fs would be required.  Note that it was necessary to increase the radial
box size to $\unit[350]{\au}$ (339 B-splines) in order to contain the double
continuum wave packet when the pulse duration exceeded $\unit[4]{fs}$.

\begin{table}[t]
    \centering 
    \begin{tabularx}{.8\columnwidth}{CCC}
        \toprule 
        Pulse duration (fs) & FWHM (fs) & Cross section ($\unit[10^{-52 }]{cm^4s}$)\\ 
        \midrule 
        2 & 0.7 & 2.6\\ 
        4 & 1.5 & 2.1\\
        5 & 1.8 & 2.0\\
        6 & 2.2 & 2.0\\ 
        $9.4^*$ & 1.8 & 2.0\\
        \bottomrule
    \end{tabularx}
    \caption{Double ionization cross sections at $\hbar\omega = \unit[51.7]{eV}$.
    ($^*$Gaussian pulse).} 
    \label{tbl:cross_section} 
\end{table}

\section{Conclusions}
In this paper we have presented a numerical method for solving the two-electron
time-dependent Schr\" odinger equation. After establishing the capability of the
method through convergence tests and accurate reproduction of known physical
quantities, we applied the method to the study of two-photon direct double
ionization of helium. Good agreement with several recently published results was
found for the total cross section. Finally, we investigated the behavior of the
cross section near the sequential threshold, where a steep increase was
observed. Calculating the value of the cross section at fixed frequency
(\unit[51.7]{eV}) for varying pulse duration, we found that the value converged
for longer pulses, and it appears that the cross section indeed exhibits a
moderate growing trend towards the threshold.

%
%
\begin{acknowledgments}
This work was supported by the Bergen Research Foundation (Norway). All
calculations were performed on the Cray XT4 (Hexagon) supercomputer installation
at Parallab, University of Bergen (Norway). The authors would like to thank
Johannes Feist, Bernard Piraux and Peter Lambropoulos for providing their
calculated cross sections in electronic form, and Henri Bachau for many useful 
discussions.
\end{acknowledgments}

\bibliography{manus}

\end{document}